\documentclass[a4paper,11pt]{article}
\pdfoutput=1

\usepackage{jinstpub}

\usepackage{siunitx}

\newcommand{\up}[1]{\textsuperscript{#1}}

\arxivnumber{2412.18021}

\title{\boldmath Calorimeter commissioning of the SuperNEMO Demonstrator}

\author[a,b]{X.~Aguerre,}
\author[c]{A.~Barabash,}
\author[d]{A.~Basharina-Freshville,}
\author[e]{M.~Bongrand,}
\author[e]{Ch.~Bourgeois,}
\author[e]{D.~Boursette,}
\author[e]{D.~Breton,}
\author[f]{R.~Breier,}
\author[g]{J.~Busto,}
\author[e]{S.~Calvez,}
\author[a]{C.~Cerna,}
\author[d]{M.~Ceschia,}
\author[a]{E.~Chauveau,}
\author[d]{L.~Dawson,}
\author[h]{D.~Duchesneau,}
\author[i]{J.J.~Evans,}
\author[c]{D.V.~Filosofov,}
\author[e]{X.~Garrido,}
\author[e]{C.~Girard-Carillo,}
\author[a]{M.~Granjon,}
\author[j]{B.~Guillon,}
\author[e]{M.~Hoballah,}
\author[k]{R.~Hod\'{a}k,}
\author[m]{J.~Horkley,}
\author[a]{A.~Huber,}
\author[d]{M.H.~Hussain,}
\author[h]{A.~Jeremie,}
\author[e]{S.~Jullian,}
\author[f]{J.~Kaizer,}
\author[c]{A.A.~Klimenko,}
\author[c]{O.~Kochetov,}
\author[k,l]{F.~Ko\v{n}a\v{r}\'{i}k,}
\author[c]{S.~Konovalov,}
\author[k,l]{T.~K\v{r}i\v{z}\'{a}k,}
\author[a]{A.~Lahaie,}
\author[n]{K.~Lang,}
\author[j]{Y.~Lemi\`ere,}
\author[b]{P.~Li,}
\author[e]{P.~Loaiza,}
\author[e]{J.~Maalmi,}
\author[k]{M.~Macko,}
\author[k]{F.~Mamedov,}
\author[a]{C.~Marquet,}
\author[j]{F.~Mauger,}
\author[k,o]{A.~Mendl,}
\author[h]{A.~Minotti,}
\author[p]{B.~Morgan,}
\author[c]{I.~Nemchenok,}
\author[q]{M.~Nomachi,}
\author[k]{V.~Palu\v{s}ov\'{a},}
\author[b]{C.~Patrick,}
\author[a]{F.~Perrot,}
\author[f,k]{M.~Petro,}
\author[a]{A.~Pin,}
\author[a]{F.~Piquemal,}
\author[f]{P.~Povinec,}
\author[b]{S.~Pratt,}
\author[n]{M.~Proga,}
\author[d]{W.S.~Quinn,}
\author[c]{A.V.~Rakhimov,}
\author[p]{Y.~Ramachers,}
\author[m]{C.~Riddle,}
\author[c]{N.I.~Rukhadze,}
\author[d]{R.~Saakyan,}
\author[n]{R.~Salazar,}
\author[r]{J.~Sedgbeer,}
\author[e]{L.~Simard,}
\author[f]{F.~\v{S}imkovic,}
\author[c]{A.A.~Smolnikov,}
\author[i,r]{S.~S\"oldner-Rembold,}
\author[k]{I.~\v{S}tekl,}
\author[s]{J.~Suhonen,}
\author[g]{H.~Tedjditi,}
\author[d]{J.~Thomas,}
\author[c]{V.~Timkin,}
\author[t,u]{V.~Tretyak,}
\author[c]{V.~Tretyak,}
\author[b]{G.~Turnbull,}
\author[e]{Y.~Vereshchaka,}
\author[d]{D.~Waters}
\author[c]{and V.~Yumatov}

\affiliation[a]{Universit\'e de Bordeaux, CNRS/IN2P3, LP2i, Bordeaux, UMR 5797, F-33170, Gradignan, France}
\affiliation[b]{University of Edinburgh, SUPA, School of Physics and Astronomy, Edinburgh, EH9 3FD, United Kingdom}
\affiliation[c]{Participant in the NEMO-3/SuperNEMO collaboration}
\affiliation[d]{University College London, London, WC1E 6BT, United Kingdom}
\affiliation[e]{Universit\'e Paris-Saclay, CNRS, IJCLab, F-91405, Orsay, France}
\affiliation[f]{Faculty of Mathematics, Physics and Informatics, Comenius University, SK-842 48, Bratislava, Slovakia}
\affiliation[g]{Aix-Marseille Universit\'e, CNRS, CPPM, F-13288 Marseille, France}
\affiliation[h]{Université de Savoie, CNRS/IN2P3, LAPP, UMR 5814, F-74941 Annecy-le-Vieux, France}
\affiliation[i]{University of Manchester, Manchester, M13 9PL, United Kingdom}
\affiliation[j]{Universit\'e de Caen Normandie, ENSICAEN, CNRS/IN2P3, LPC Caen, UMR6534, F-14000, Caen, France}
\affiliation[k]{Institute of Experimental and Applied Physics, Czech Technical University in Prague, CZ-11000 Prague, Czech Republic}
\affiliation[l]{Faculty of Nuclear Sciences and Physical Engineering, Czech Technical University in Prague, Brehova, 7 115, Czech Republic}
\affiliation[m]{Idaho National Laboratory, Idaho Falls, ID 83415, USA}
\affiliation[n]{University of Texas at Austin Department of Physics,
Austin, TX 78712, USA}
\affiliation[o]{Faculty of Mathematics and Physics, Charles University, CZ-12116 Prague, Czech Republic}
\affiliation[p]{University of Warwick, Coventry, CV4 7AL, United Kingdom}
\affiliation[q]{Osaka University, 1-1 Machikaneyama Toyonaka, Osaka 560-0043, Japan}
\affiliation[r]{Imperial College London, London, SW7 2BZ, United Kingdom}
\affiliation[s]{Department of Physics, University of Jyv\"askyl\"a, Jyv\"askyl\"a, Finland}
\affiliation[t]{Institute for Nuclear Research of NASU, Kyiv, 03028, Ukraine}
\affiliation[u]{INFN, Laboratori Nazionali del Gran Sasso, 67100, Assergi (AQ), Italy}

\collaboration{SuperNEMO Collaboration:}

\emailAdd{SUPERNEMO-L@IN2P3.FR}

\abstract{The SuperNEMO experiment is searching for neutrinoless double-beta-decay of \up{82}Se, with the unique combination of a tracking
detector and a segmented calorimeter. This feature allows us to detect
the two electrons emitted in the decay and measure their individual
energies and angular distribution. The SuperNEMO Demonstrator's
calorimeter consists of 712  plastic scintillator blocks read out by
large PMTs. Having constructed the calorimeter underground, we performed
its first commissioning using $\gamma$-rays from calibration
sources or from the ambient radioactivity background. This article
presents quality assurance tests of the SuperNEMO Demonstrator's
calorimeter, and its first time and energy calibrations with $\gamma$-rays, along with the
associated methods. A time alignment of about 120~\si{\pico\second} and a time resolution around 615~\si{\pico\second} have been achieved. Concerning the energy, an alignment of 7.5\% has been obtained. These results will be further improved when associating the tracking detector and detecting electrons from calibration sources.}

\graphicspath{{figures/}}

\begin{document}
\maketitle
\flushbottom

\section*{Introduction}
\addcontentsline{toc}{section}{Introduction}

SuperNEMO is a double-beta-decay (DBD) experiment, designed to look for
the hypothesized lepton-number-violating process of neutrinoless
double-beta-decay (0$\nu\beta\beta$) and the Majorana nature of the
neutrino.

SuperNEMO's tracker-calorimeter design, based on the technology of the NEMO-3 detector
\cite{Arnold:2004xq}, allows for the study of different isotopes and
tracking of the outgoing MeV-energy-scale electrons. This ability
provides the means to discriminate between different underlying
mechanisms of 0$\nu\beta\beta$  by measuring the decay half-life and the
electron angular and individual energy distributions \cite{Arnold:2010tu}.
It is also well suited for precision studies of the Standard Model
two-neutrino double-beta-decay (2$\nu\beta\beta$) which is present in
all 0$\nu\beta\beta$ candidate isotopes \cite{NEMO:2008kpp,
NEMO-3:2009fxe, NEMO-3:2011wyg, NEMO-3:2016mvr, NEMO-3:2016zfx,
Arnold:2018tmo, Arnold:2019mo100bb2n}. In particular, the measurement of
the full topology of 2$\nu \beta \beta$ events could constrain the
quenching of the axial-vector coupling constant ($g_A$)
\cite{simko:2018}. Adding the detection of $\gamma$-rays to the DBD
events, NEMO technology can also precisely study decays to excited
states of the daughter nuclei \cite{NEMO:2006smm, NEMO-3:2020ssy,
NEMO-3:2022nvb}.

\begin{figure}[htbp]
\centering
\includegraphics[width=\textwidth]{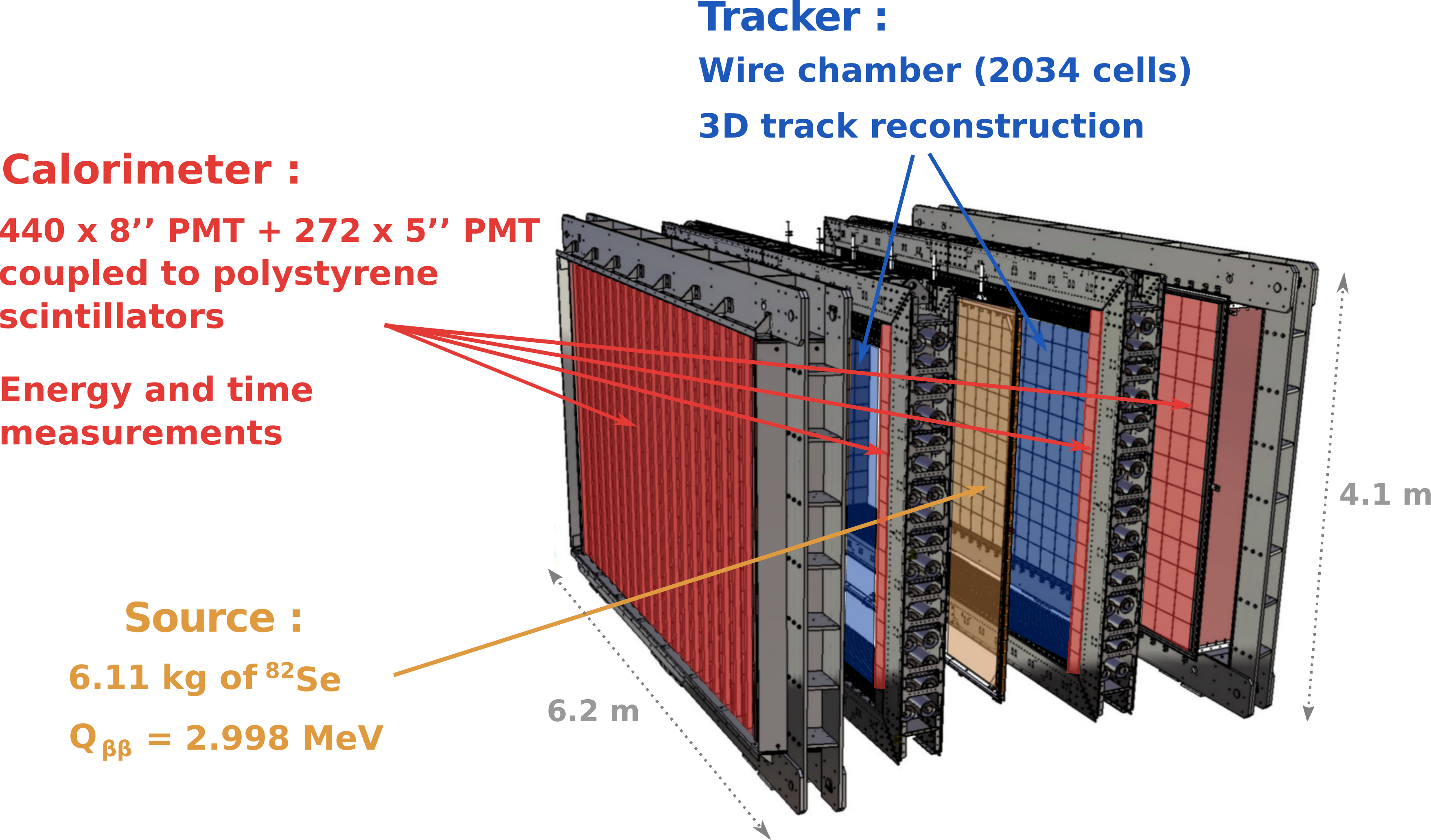}
\caption{\label{fig:demonstrator-design} Split view of the sub-detectors
of the SuperNEMO Demonstrator module.}
\end{figure}

The basic unit of SuperNEMO is a module, as shown in
figure~\ref{fig:demonstrator-design}. The modular design allows the
detector's size to be increased by adding identical modules to the
detector. The first module, named the \textit{SuperNEMO Demonstrator},
is currently undergoing the final stages of installation and
commissioning at the \textit{Laboratoire Souterrain de Modane} (LSM), in
France.

\begin{figure}[htbp]
\centering
\includegraphics[height=0.3\textheight]{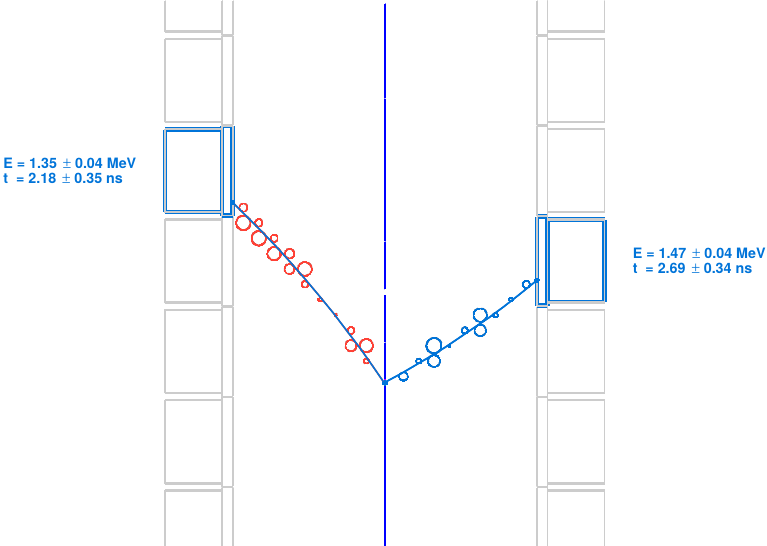}
\caption{\label{fig:2e-event} Top view of a simulated (with GEANT-4) DBD
event illustrating the detection principle of the SuperNEMO detector.
The decay electrons are emitted from the vertical isotopic source foils
(in blue). The 3D tracks of the electrons are fitted from the drift
cells signals (red and blue circles) of the tracking detector. Arrival
time and energy of each electron can then be measured by the segmented
calorimeter (grey rectangles highlighted in blue).}
\end{figure}

To prevent radioactive contamination from dust, a dedicated clean tent
was built prior to the Demonstrator's construction. The subdetectors of
the Demonstrator were successively integrated inside this clean tent.
 From the center to the edges, the detector consists of (see also
figure~\ref{fig:2e-event}):

\begin{itemize}\setlength{\itemsep}{0pt}
\item \textbf{A source frame} containing 34 isotopically-enriched DBD
source foils, for a total of 6.11$\,$kg of \up{82}Se. The source foils
are 2.7 m long with a width of around 135$\,$\si{\mm} and a thickness of
around 300$\,$\si{\micro\meter}. The foils are made of purified selenium
powder mixed with radiopure polyvinyl alcohol glue supported by two
12$\,$\si{\micro\meter}-thick mylar films \cite{Jeremie:20177b, SuperNEMO:source}. At
each end of the plane of selenium sources, one pure copper strip has
been added as a control, to monitor the external - to the DBD source -
background, in a radiopure, non-double-beta-decay sample.
\item \textbf{Two tracking detectors}, made of 2034 vertical drift
cells, surround the source frame. The drift cells are used to
reconstruct the tracks of the $\beta$-particles emitted by the sources
in three dimensions. The dimensions of one tracking volume are 3.1$\,$m
tall, 5.0$\,$m wide and 0.44$\,$m deep. The total tracking volume thus
corresponds to about 15.4$\,$\si{\cubic\meter}. The tracking detector
gas is a mixture of 95\% helium, 1\% argon and 4\% ethanol.
\item \textbf{A segmented calorimeter} consisting of plastic scintillators coupled
to photomultiplier tubes (PMTs) to measure the deposited energy and the
time of flight of particles coming from the double-beta sources or from
backgrounds. Calorimeter modules surround the edges of the tracker
frames, and two main calorimeter walls close the detector on each side
of the tracker, parallel to the source frame. To improve the
tracking-gas tightness and reduce radon penetration, a
25$\,$\si{\micro\meter} polyamide film separates the tracker and the
calorimeter.
\item \textbf{The calibration tools} are comprised of an automated
deployment system of radioactive sources at the top of the central frame and a
light-injection system for the calorimeter blocks. The deployment system
allows insertion of 42 low-activity (120-145 Bq) sources of \up{207}Bi
\cite{SuperNEMO:2021hqx} between the \up{82}Se foils, from six vessels
at the top of the source frame. The calibration sources emit conversion
electrons of around 500 keV and 1 MeV. They are used for regular absolute
energy calibrations, on a weekly basis. In addition, a light injection
system using pulsed ultra-violet (385 nm) LEDs and optical fibers, to
reach the plastic scintillators, allows more regular surveys of the
stability of the PMTs, several times per day. This system is accompanied
by five additional reference calorimeter blocks permanently equipped
with electron (\up{207}Bi) and alpha (\up{241}Am) radioactive sources.
These are used to compare the LED pulses to calibration-source signals
to monitor the LEDs.
\item \textbf{A copper magnetic coil} with iron return field plates
surrounds the whole detector. The 25 G magnetic field will improve the
identification of the DBD electrons and enhance the background rejection
(for example $e^+e^-$ pairs or crossing electrons). The PMTs are
protected from the magnetic field by pure iron shields.
\item \textbf{A gas-tight anti-radon tent} flushed with deradonized air
($\sim$10 \si{\milli\becquerel\per\cubic\meter}) produced at LSM, using
an anti-radon facility, prevents radon in the laboratory air from
entering the detector.
\item \textbf{Two shielding layers}, finally, enclose the detector to
reduce the background from natural radioactivity of the underground
laboratory: 18$\,$cm thick iron shielding to reduce the $\gamma$~flux
and 50 cm water or 24 cm polyethylene shielding to reduce the neutron flux.
\end{itemize}

The detection principle of SuperNEMO is illustrated in figure~\ref{fig:2e-event}, with the example of a simulated two-electron
event.\\

At the time of writing of this article, the detector itself is fully
assembled at the LSM, including the magnetic coil and the anti-radon tent. The
final shielding is being integrated. Once the SuperNEMO calorimeter was
fully installed, with its cabling, high-voltage (HV) and front-end
electronics (FEE) integrated, an early
commissioning of the calorimeter was carried out. This involved measurements of
$\gamma$-rays before the tracking detector was operational. This
article focuses on the main-wall calorimeter, describing the
commissioning methods, the initial calibrations, and presenting the
preliminary performance of the calorimeter.\\

In this article, we  present the SuperNEMO calorimeter design in
section~\ref{sec:supernemo-calorimeter}. We then give an overview of the
preliminary tests concerning the cabling, the electronics and the PMTs
in section~\ref{sec:calorimeter-preliminary-tests}. The two last
sections present the timing (section~\ref{sec:time-calibration}) and
energy (section~\ref{sec:energy-calibration}) calibration, and
performance of the calorimeter.

%-----------------------------%

\section{The SuperNEMO calorimeter}
\label{sec:supernemo-calorimeter}

The SuperNEMO calorimeter is primarily dedicated to the detection of the
DBD electrons emitted by the $\beta\beta$-sources. However, the
detection of $\gamma$-rays is also of major importance for the
study of DBD to excited states and to the identification
of background events. The SuperNEMO calorimeter must therefore be
efficient in gamma detection and offer a full solid-angle coverage. In
total, the SuperNEMO calorimeter is segmented into 712 optical modules
(OMs) to detect particles individually. The ability to measure the
energy of each particle separately is a unique feature in the field of DBD experiments. An illustration of the measurement of the separate
electron energies is illustrated in figure~\ref{fig:2e-event}.

\begin{figure}[!htb]
\centering
\includegraphics[height=0.28\textheight]{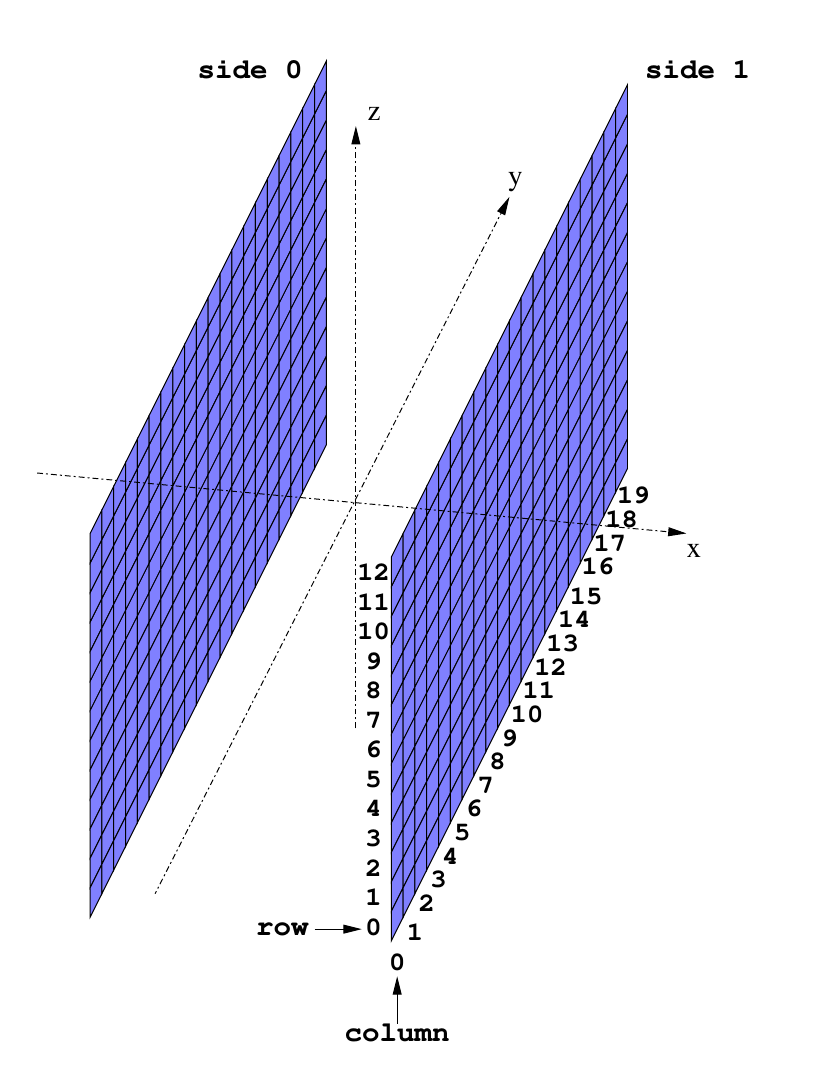}
\includegraphics[height=0.24\textheight]{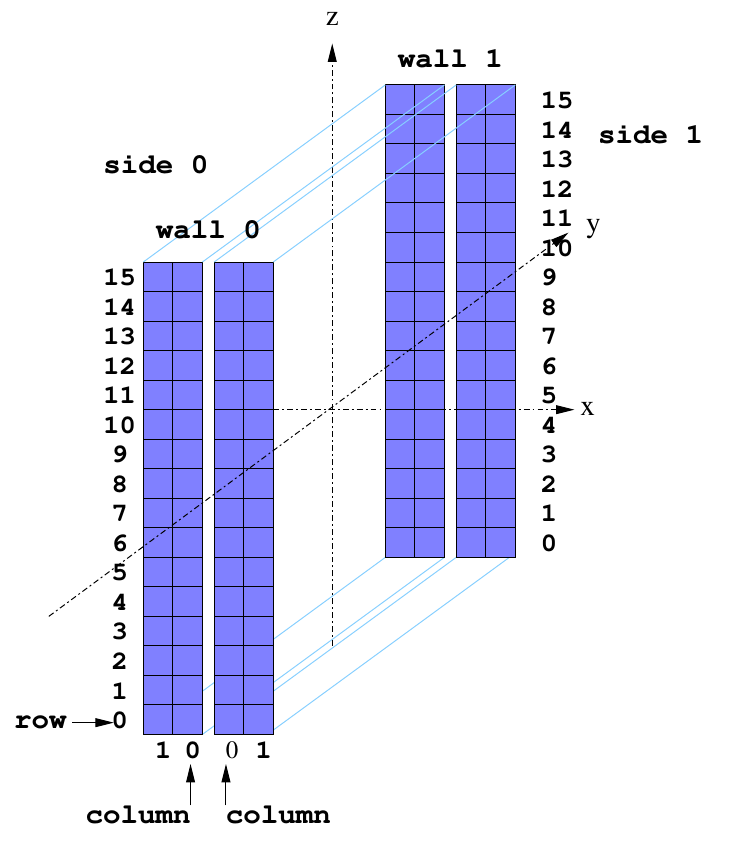}
\includegraphics[height=0.23\textheight]{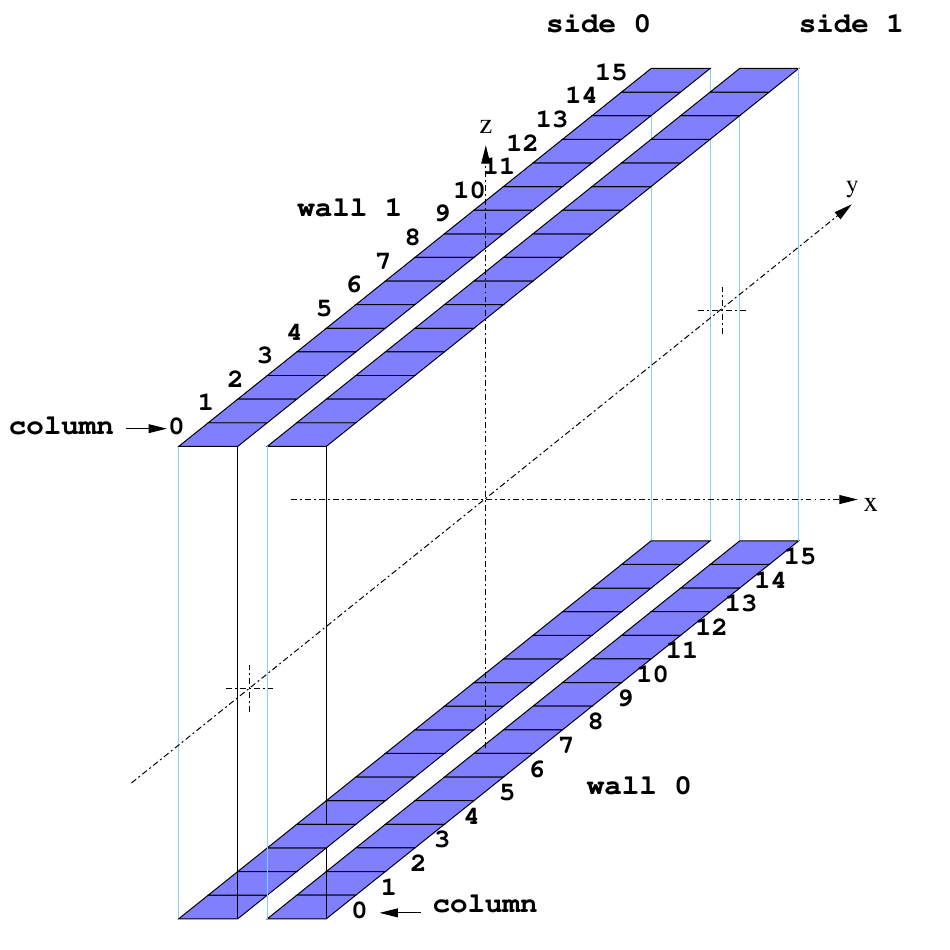}
\caption{Illustration of the SuperNEMO calorimeter segmentation with
M-wall (left), X-wall (middle) and G-veto (right) optical modules. The
two sides of the detector are sometimes referred to as \textit{French}
(side 1) or \textit{Italian} (side 0) sides, corresponding to which
country of the road tunnel they are facing.}
\label{fig:calo-labelling}
\end{figure}

Each of the two calorimeter main walls, {\em M-wall}, are segmented into
$20\times13=260$ OMs. Each tracking detector has $2\times16=32$ cross-wall OMs, {\em
X-wall}, on each edge, and 16 gamma-veto OMs, {\em G-veto}, on the top
and the bottom. The positions of these three categories of OMs are
presented in figure~\ref{fig:calo-labelling}.

\subsection{The optical modules}

\begin{figure}[htbp]
\centering
\includegraphics[width=\textwidth]{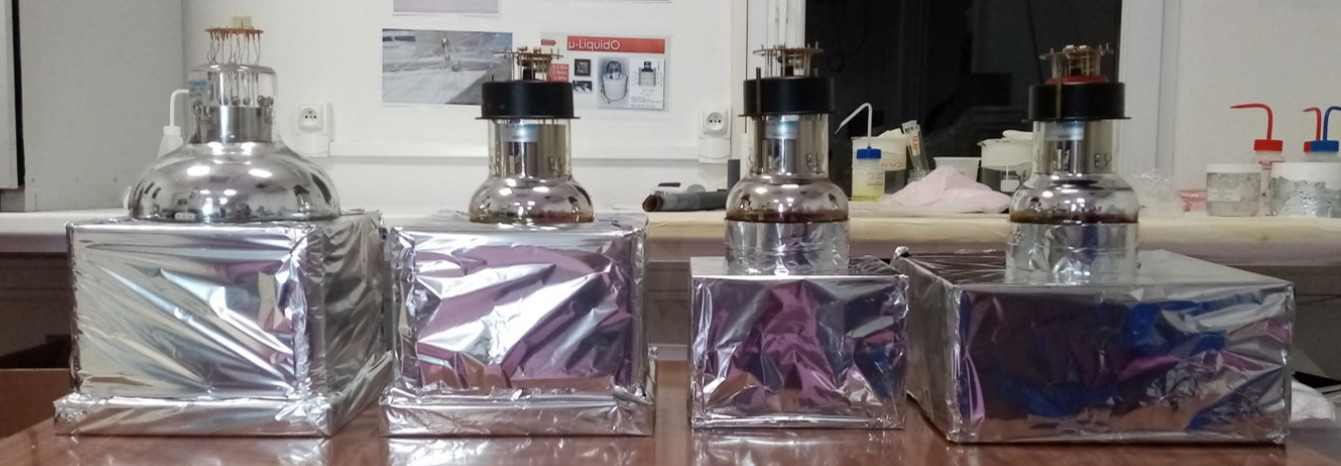}
\caption{\label{fig:oms-picture} Picture of the four types of optical
modules used for the SuperNEMO calorimeter during assembly. From left to
right: two M-wall OMs with 8- and 5-inch PMTs, one X-wall OM and one
G-veto OM with 5-inch PMTs and PMMA light-guide.  Notice the wider step in the M-Wall scintillator block’s shape at the bottom of the corresponding OM. Scintillators and
light-guides are wrapped with aluminized mylar.}
\end{figure}

One OM consists of an assembly of an organic scintillator and a
photomultiplier tube glued with RTV-615 optical glue, as can be seen in figure~\ref{fig:oms-picture}. The scintillator
is a polystyrene (PS)-based organic scintillator doped with 0.05$\,$\%
of POPOP (1.4-bis(5-phenyloxazol-2-yl)benzene) and 1.5$\,$\% of pTP
(p-Terphenyl) wavelength shifters \cite{Barabash:2017sxf,
10.1063/1.4928013}. Two productions of PS scintillators have been used:
a standard production and an enhanced one, which provides a better light
yield and a higher transparency. Eight scintillator blocks of
polyvinyltoluene (PVT), with even higher light yield, have also been
included. The scintillator is wrapped with Teflon on the four lateral
sides, and two layers of 6$\,$\si{\micro\meter} aluminized mylar around
all six faces (except the surface in direct contact with the PMT) to
improve the light collection towards the PMT. The thickness of the front
face of the scintillator wrapping is minimized to reduce energy losses
for the entering $\beta$-particles. Two types of PMTs have been used
with the scintillators: new 8-inch Hamamatsu R5912-MOD; and 5-inch
Hamamatsu R6594, refurbished from the previous NEMO-3 experiment
\cite{Arnold:2004xq}.

Plastic scintillators have been chosen for their excellent radiopurity,
high light yield, fast time response, and low back-scattering
probability for incident electrons. To investigate DBD to excited states
or reject and study backgrounds, the SuperNEMO calorimeter has also been
designed to detect $\gamma$-particles. The depth of the scintillators
has thus been increased to almost 20$\,$cm, while less than 3$\,$cm
would be enough to contain the DBD electrons.

As can be seen in figure~\ref{fig:oms-picture}, four types of optical
modules have been used for the SuperNEMO calorimeter. 

The configuration of OMs in the SuperNEMO calorimeter has been chosen to optimise our energy resolution when detecting DBD electrons, which will be mostly detected on the two main calorimeter
walls, while also fitting the constraints of the tracker’s mechanical design. The M-wall OMs are made of the best plastic scintillators with
8-inch PMTs, to achieve 8 \% FWHM energy resolution at 1$\,$MeV, except
for the top and bottom rows, where 5-inch PMTs are used, since the
end-caps of the tracking cells prevent the $\beta$-particles emitted
from the DBD sources from being detected there. To reach the best energy
resolution and light-collection uniformity possible,
large-photocathode-area PMTs are glued directly to these M-wall scintillators,
without interface light guides. These scintillators also have a 31 mm step, providing a larger surface area
compared to the rest of the block (steps visible in
figure~\ref{fig:oms-picture}) to maximize the detection surface and to
leave space for the magnetic shield surrounding all the M-wall OMs. On
the small sides of the trackers, parallel to the tracking cells, the
X-walls are made of smaller scintillators coupled to 5-inch PMTs with a
PMMA light-guide. A cylindrical mu-metal magnetic shield covers the PMT
and its light guide; as the PMT is small, it is unnecessary to shield
the scintillator. These X-wall OMs can also detect DBD
particles from the source foils, but with a poorer energy resolution of
12$\,$\% FWHM at 1 MeV. A similar design is used for the G-veto OMs,
which are installed on the top and bottom of the trackers, but with
larger scintillator blocks. Similarly to the top and the bottom OMs of
the M-wall, these OMs will not detect the $\beta$-particles of the
sources because they are placed above or below the end-cap of the
tracking cells. The constraint on the energy resolution can be relaxed
and they achieve 16 $\,$\% FWHM at 1$\,$MeV. More details about the
development of these optical modules and the radiopurity budget can be
found in \cite{Barabash:2017sxf}.

To ensure the radiopurity of the SuperNEMO calorimeter, the components
of the OMs have been carefully selected using high-purity germanium
(HPGe) detectors. The glass of the PMTs (650 g) and their electrical
insulators (25 g) represent the major contribution to the radionuclide activities of
the OMs. For the 8-inch Hamamatsu PMTs, contamination levels of
0.95$\,$Bq/PMT in \up{40}K, 0.62$\,$Bq/PMT in \up{214}Bi and
0.26$\,$Bq/PMT in \up{208}Tl have been measured.  Only upper limits of a
few mBq/kg for the three main radionuclides have been set on the
radiopurity of the plastic scintillators.

\subsection{The main wall calorimeter frame}

The dimensions of the calorimeter frame have been driven by the source
surface to offer an active area of about 5$\times$3~\si{\square\meter}.
It would be difficult to transport a detector of this size to the
Laboratoire Souterrain de Modane (LSM) in one piece. Furthermore, these
dimensions could not fit through the LSM's entrance from the tunnel. The
frame was therefore designed as an assembly of four beams to be
assembled underground, and to be populated with the optical modules
onsite. The structure of the calorimeter frame has to support the weight
of 260 OMs that make up each main wall, each weighting around 25$\,$\si{\kilo\gram}, and totaling
about 6.5$\,$tons. The four beams consist of reinforced structures made
from 30$\,$mm-thick pure iron plates. This material has been selected
for mechanical considerations, and for its very good radiopurity.

\begin{figure}[!htb]
\centering
\includegraphics[width=0.9\textwidth]{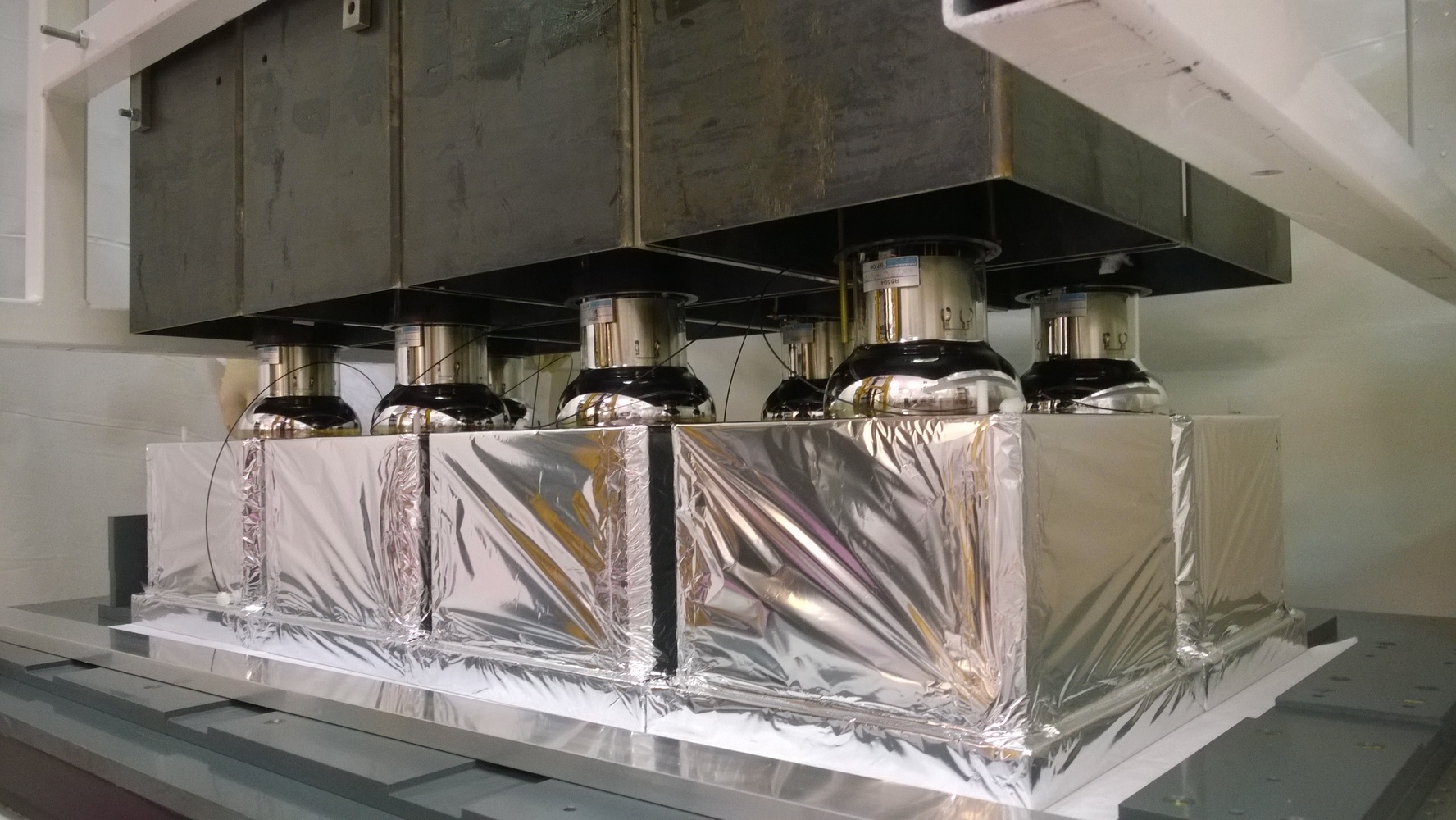}
\caption{Picture of a calobrick of 2$\times$4 OMs during insertion of
the pure iron magnetic shields.}
\label{fig:calobrick}
\end{figure}

Because of the fragility of the scintillator wrapping, and to speed up
the integration process underground, the OMs were packed into so-called
{\em calobricks}. These calobricks were horizontal assemblies of
4$\times$2 or 4$\times$1 OMs, as shown in figure~\ref{fig:calobrick}.
The support structure of the calobricks relies on the iron magnetic
shields, which are screwed together with radiopure brass bolts and
separated by PMMA spacers. The OMs are attached to the transverse plate of the magnetic shield by nylon screws. The magnetic shields have been developed to
protect the PMTs from the 25 G magnetic field, needed in the tracker for electron-positron separation via the curvature of their tracks. They consist of
3 mm-thick ARMCO pure iron plates, cut and soldered by laser, to prevent
contamination. This material was preferred to mu-metal because of its
better radiopurity (only upper limits at 90\% C.L. have been measured at
the level of 5.7$\,$mBq/kg in \up{40}K, 1.1$\,$mBq/kg in \up{214}Bi and
0.48$\,$mBq/kg in \up{208}Tl). Special annealing treatment was used to
improve the magnetic properties of the shields after production. The
shields cover a large part of the scintillator because the need to go
beyond the PMT photocathode has been demonstrated, to reduce the
penetrating magnetic field. There is, however, a residual magnetic field
that still penetrates the PMT, of the order of 1 G. In the end, only a
few percent loss has been observed on the PMT charge, which could be
compensated for by a moderate increase in the high voltage, on the order
of 20 V \cite{calvez:tel-01632815}.

\subsection{The calorimeter cabling}

\begin{figure}[!htb]
\centering
\includegraphics[height=0.6\textheight]{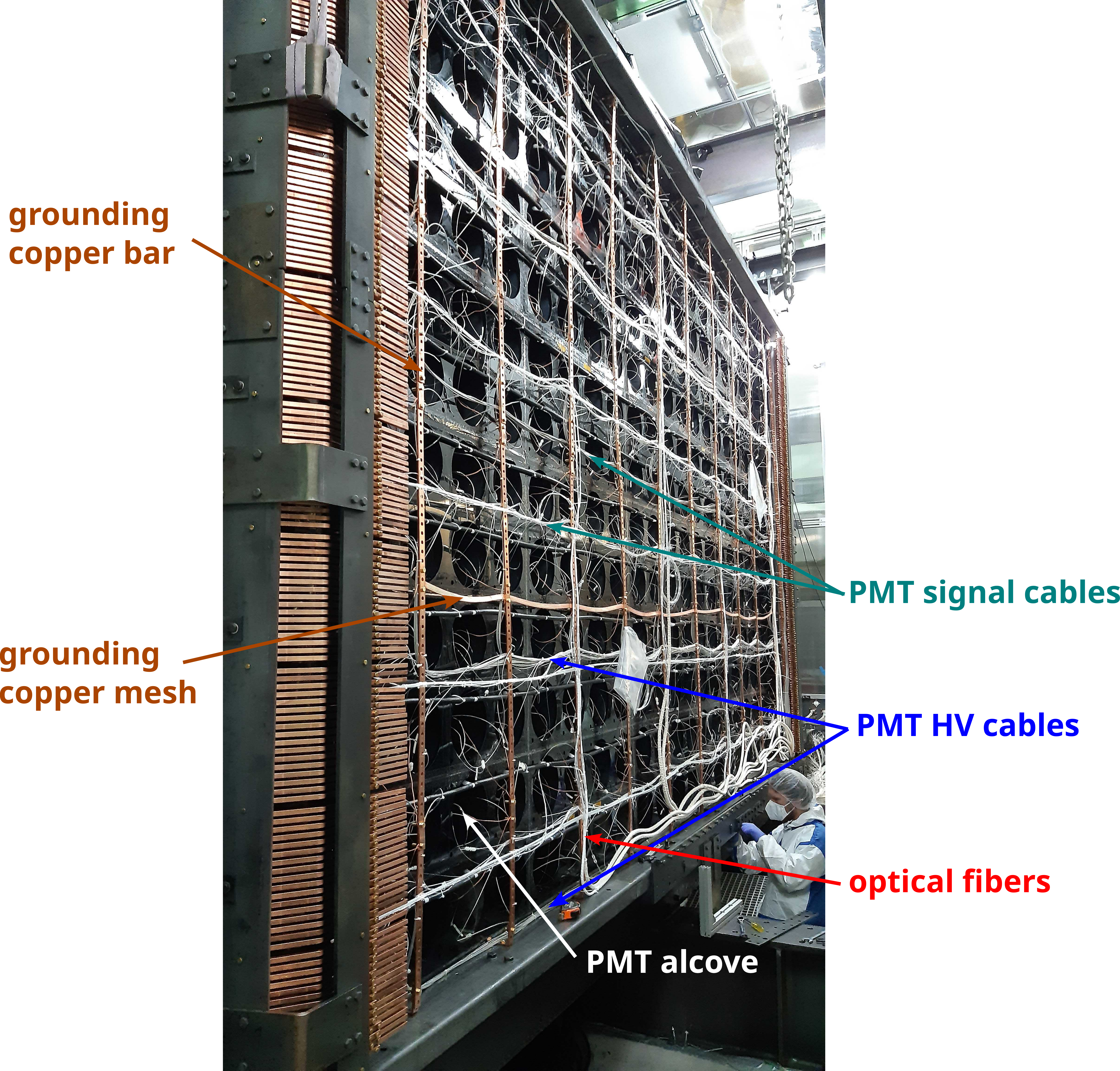}
\caption{Picture of the back of one calorimeter main wall showing the
horizontal routing of the signal and HV cables to the PMT dividers in
the alcoves, the vertical grounding copper bars, the large and small
copper braids and the optical fibers of the light injection system in
their white jackets.}
\label{fig:calo-cabling}
\end{figure}

The SuperNEMO detector is protected by an air-tight anti-radon tent
flushed with radon-free air. The calorimeter electronics are outside of
this tent;  the HV and signal cables, connecting the PMTs to the
electronics, have to go through the tent while maintaining the gas
tightness. For this reason, different sets of cables are used inside
(see figure~\ref{fig:calo-cabling}) and outside of the anti-radon tent.
Internal and external cables are connected together at the patch panels
on the anti-radon tent with dedicated connectors. The patch panels are
made of drilled pure-iron plates to accept, from inside, fixed female
connectors and, from outside, removable cables with their male connectors.

\subsubsection*{Signal cabling}

The signal cables of the SuperNEMO calorimeter have been selected for its
radiopurity and compatibility with the front-end electronics. It is a
coaxial cable with the reference M17/93 RG-178 from the Axon company,
with a transparent sheath that provides a better radiopurity. Only the
contamination in \up{40}K has been measured, at 25$\pm$4
\si{\milli\becquerel\per\kilo\gram}; 90\% C.L. upper limits have been
set at 1.1$\,$\si{\milli\becquerel\per\kilo\gram} in \up{214}Bi and
0.36$\,$\si{\milli\becquerel\per\kilo\gram} in \up{208}Tl. In order to
reduce the amount of cables inside the anti-radon tent, the cables have
been cut to the required lengths to reach each PMT by first a vertical
routing to the OM's row, and then a horizontal routing to the PMT's
location. The cables are fastened to vertical copper bars attached to
the calorimeter's pure-iron frame. This scheme allows the mapping of the
OMs in the detector to be reflected at the patch panel and at the
front-end electronics boards (see
section~\ref{sec:calorimeter-electronics}). The separation between two adjacent OMs are 25$\,$\si{\cm}
between OMs in both directions. This results in internal signal cables
with lengths ranging from 3.25$\,$\si{\m} to 11$\,$\si{\m}. The
difference in cable lengths is compensated for offline by time alignment
of all the channels, as described in section~\ref{sec:time-calibration}.
The cables are connected to the PMT dividers using two female Souriau
pins (SC20WL3S25) for the inner connector. The other end of the internal
signal cable has a female MCX coaxial connector (MP-27-10M TGG) straight
bulkhead jack to be fixed on the patch-panel plate.

Given the relatively small dimensions of the patch panel and the
compactness of the front-end electronics, all the external cables have
been cut to the same length of 7$\,$\si{\m}. For simplicity, the same
Axon cable has been used as for the internal cables. On the SuperNEMO
front-end boards (section~\ref{sec:calorimeter-electronics}), a female
MCX 50 Ohms coaxial connector is used, similar to that used at the patch
panel for the internal cable. Thus, the external signal cables have male coaxial MCX connectors from Radiall on both ends.

\subsubsection*{HV cabling}

For the same reasons as for the signal cables, the HV cables are divided
into internal and external cables at the anti-radon tent. Power for the
SuperNEMO calorimeter is supplied by three CAEN SY4527 units. The crates
are populated with 32-channel A1536 HV boards. These boards come with
52-pin connectors from Radiall. To fit with this connector, an HV cable
produced by CERN was selected for the external part. It consists of a 37
multi-cables bundled in a red jacket. In this cable, 32 channels are used
to supply HV to the individual PMTs, and 5 channels are used to provide
grounding. The connectors on this external HV cable are a Radiall
connector (691802002) at the electronics side, and Redel LEMO 51-pin
straight plug (SAG.H51.LLZBG) at the patch-panel side.

Inside the anti-radon tent, the HV cables become individual coaxial
cables. The connector at the patch panel is a fixed-socket Redel LEMO
(SLG.H51.LLZG) corresponding to the female external one. An in-house
circuit has been designed to merge the grounds at this level, and
distribute all the HV to the dedicated pins of the connector to the
individual internal cables. The internal cable is an Axon AK4902A
coaxial HV cable, also selected for its good radiopurity. Contamination
levels in \up{40}K and \up{214}Bi have been measured at
23$\pm$4$\,$\si{\milli\becquerel\per\kilo\gram} and
1.1$\pm$0.3$\,$\si{\milli\becquerel\per\kilo\gram} respectively, while a
90\% C.L. upper limit has been set to
0.3$\,$\si{\milli\becquerel\per\kilo\gram} in \up{208}Tl. Similarly to
the signal cable, the external HV cable can be disconnected from the
outside, while the internal one is fixed to the patch panel. The
internal cables are connected to the PMT dividers using two Souriau pins
(one male SM20WL3S26 for the core, and one female SC20WL3S25 for the
ground mesh, to prevent misconnection). A small copper braid has also
been added to connect the PMT divider to the mechanical grounding of the
calorimeter wall.

The routing scheme of the internal cables on the calorimeter is similar
to the signal cables, but with a different pattern, to fit the mapping
between OMs and the channels on the HV boards: see
section~\ref{sec:calorimeter-electronics}. We have also avoided routing
signal and HV cables together, to reduce possible noise or cross-talk.

\subsection{Electrical grounding}

In order to prevent noise being induced on the signal channels and
electric charge accumulation in the detector, special care has been
taken to ground the SuperNEMO calorimeter. Two types of copper braids
have been deployed to ensure this grounding. Firstly, a large copper
braid (10$\,$\si{\milli\meter\squared}) is routed all along the
calorimeter walls (see figure~\ref{fig:calo-cabling}) and connected to
the vertical copper bars also used to support the PMT cables. The
connection is made by pinching the copper braid with a copper plate on
the vertical bars. To connect each PMT divider to ground, a second
copper braid has been added (1.5$\,$\si{\milli\meter\squared}). It is
soldered to the naked ground mesh of each HV cable and then pinched to
the large copper braid with small copper plates.

The whole detector and its electronics are installed on brass plates
that ensure the overall grounding of the experiment. These brass plates
are connected to the general grounding of the LSM. All the apparatus and
sub-detectors are grounded to these plates through copper plates or braids.

\subsection{Calorimeter electronics}
\label{sec:calorimeter-electronics}

The SuperNEMO electronics are installed fully outside the detector
and its shielding. All the electronic components are grouped inside six
electronics racks. Two racks are dedicated to the calibration systems,
two others to the calorimeter, and two more to the tracker. CAEN
systems provide the high voltage (HV) to the PMTs. Custom-made front-end
boards (FEB) digitize the signals \cite{Breton20052853}. These FEBs are
inserted inside {\it versa module eurocard} (VME) crates with a custom
backplane developed for the experiment. The FEBs are handled by a
control board (CB) in each VME crate. The CBs ensure the synchronization
of the FEBs, exchange the trigger decisions, and concentrate and
transmit the signals registered by the FEBs. Both calorimeter and
tracker electronics are managed by a single trigger board (TB) connected
to all the CBs and the DAQ. The TB distributes the 40 MHz clock (25 ns
clock tick) and takes the triggering decision, which is configurable. A
complete description of the SuperNEMO electronics can be found in
\cite{oliviero:tel-01929176}. One of the racks also includes the
computers for DAQ, data storage, and a switch for network distribution.
The collected data is transferred daily to CC-IN2P3\footnote{\href{https://cc.in2p3.fr/en/}{https://cc.in2p3.fr/en/}} for storage,
processing and analysis.

The SuperNEMO trigger system is designed to select only events of
physical interest and reject spurious events (self-triggering of the
drift cells or PMTs) in order to reduce the global acquisition rate.  At
each clock tick (25 ns), the trigger system collects and merges the 712
calorimeter channels' statuses (triggered/not triggered) based on an
optimized signal-amplitude threshold per channel. A column of PMTs (13
PMTs for main walls, or 16 PMTs for X-walls and G-vetos) is connected to
a single calorimeter FEB. Each FEB builds a trigger-primitive bitset
which is sent to the dedicated CB each 25$\,$ns. Constrained by the
bandwidth from the FEBs to the CB, this bitset is limited to 5 bits. The
trigger-primitive bitset is composed of a threshold multiplicity at the
level of the FEB. At each clock tick, the CBs collect the
trigger-primitive bitset from the connected FEBs (20 $\times$ 5 bits
from each main wall and 12 $\times$ 5 bits from all of the X-walls and
G-vetos). Each CB produces a crate trigger bitset of 18 bits. The bitset
is composed of: a threshold multiplicity at the level of the crate (from
0 to $>$3) and 10 bits called the \textit{zoning word}. Each bit of this
zoning word can be activated in case of hits in the corresponding
geographical location of the detector. Three crate trigger bitsets from
the calorimeter crates are uploaded to the TB. This strategy enables a
check for coincidence between tracks from the tracker and hits from
PMTs, in time and space.

The PMT pulses are digitized in the FEBs at a sampling rate of 2.56 GS/s
over 1024 samples, providing a total time window of 400 ns. The
calorimeter data are described by a header and the data itself. The
header is composed of: the trigger ID, the electronic channel address
and the timestamp, which is the absolute time of the experiment tagging
the last sample encoded by the 48-bit TDC counter provided with a 160
MHz clock by the SAMLONG chip (390.625 ps steps), which is the heart of the calorimeter FEBs. The data contain the
metadata computed by the SAMLONG chip \cite{Breton:2012rj} and the
digitized samples of a pair of detector channels connected to the chip.
The recorded digitized waveforms are 1024 values of ADC encoded in 12
bits, corresponding to the 1 V input range of the chip (0.25 mV steps).
The metadata includes a raw estimation of the signal baseline, a charge,
a peak amplitude and values at specific times along the waveform.

\begin{figure}[htb]
\centering
\includegraphics[width=0.7\textwidth]{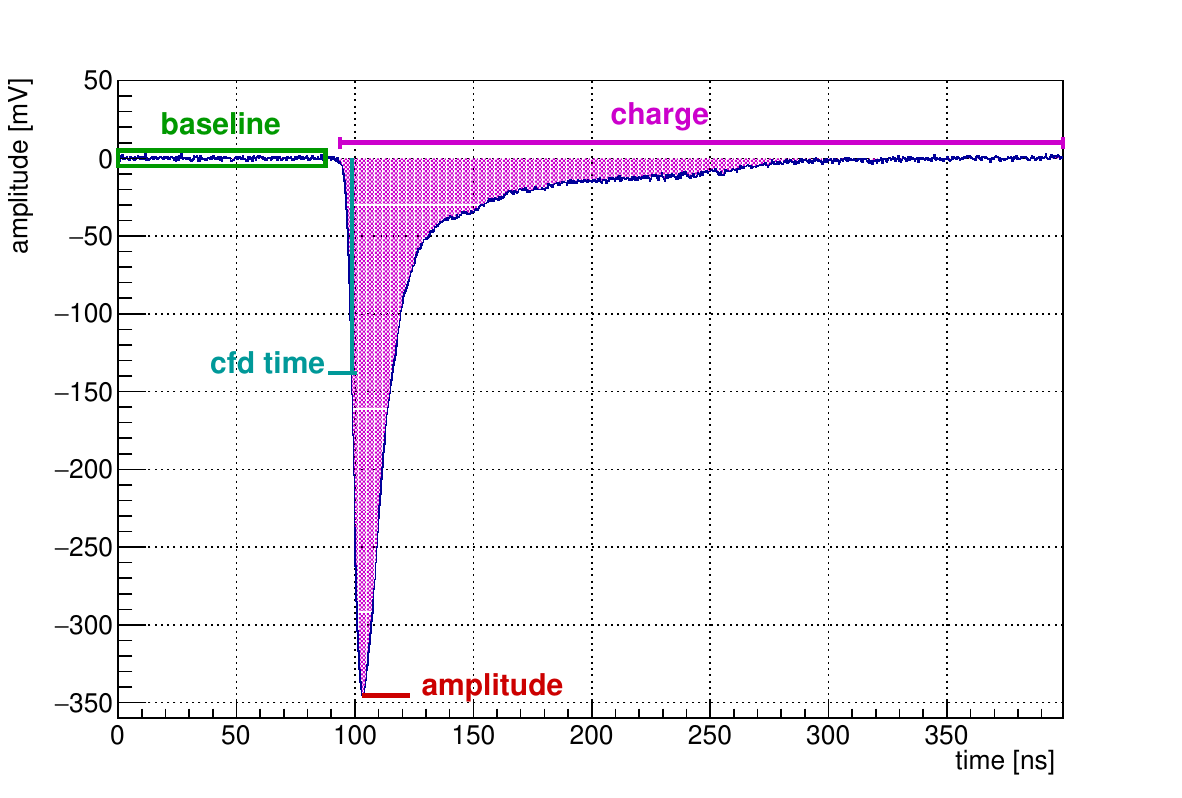}
\caption{\label{fig:pmt-pulse} Example of an 8" PMT pulse digitized by
the SuperNEMO front-end electronics at 2.56$\,$GS/s, illustrating the
offline reconstructed parameters for the analyses. The baseline is
computed by averaging the noise over the samples before the pulse
\cite{tedjd2021}. The pulse amplitude corresponds to the minimal sample
of the waveform. The time of the pulse is computed by a constant
fraction discriminator (CFD) at 40\% of the amplitude. Finally, the
charge is the integration of the pulse over the time window after the
baseline. All the parameters for these computations can be adjusted.}
\end{figure}

In order to refine the calculation of these parameters for the
determination of the physics variables of interest in each event,
offline pulse-shape analysis algorithms have been developed on the SuperNEMO PMT
signals. Four parameters of major interest are
computed for each PMT pulse: \textit{baseline, amplitude, time} and
\textit{charge}. An illustration of the computation of these parameters
on an 8" PMT signal of the SuperNEMO calorimeter is presented on
figure~\ref{fig:pmt-pulse}. The use of a constant fraction discriminator
(CFD) is independent of the amplitude of the pulses in order to improve
the precision on the time measurement, compared to NEMO-3 where
time-amplitude corrections were used.

%-----------------------------%

\section{Calorimeter preliminary tests}
\label{sec:calorimeter-preliminary-tests}

After the detector closure and the cabling operations on the
calorimeter, the front-end electronics were installed,  so that
calorimeter-commissioning data could be taken. The first stage of the
calorimeter commissioning consisted of testing the whole cabling
network, before investigating the PMT signals and determining the
performance of the calorimeter.

\subsection{HV cabling tests}

The test of the HV cables was performed by a basic method:
progressively  increasing  the voltages applied to each PMT to their
nominal values (around 1500 V). After fixing minor issues, the HV
delivery system of the SuperNEMO calorimeter was fully operational.

This required the detector to be covered with black plastic light
protection and for the the light to be switched off in the LSM, since
the detector was not light-tight at this stage of integration. In order
to avoid this operation, and to preserve access to each OM, an
alternative method was used to test the signal cabling, as described below.

\subsection{Signal cabling quality with reflectometry measurements}
\label{ssec:reflectometry-test}

The first quality-assurance tests performed on the signal cabling were
possible without switching on the PMTs,  thus avoiding the need to cover
the whole detector for light tightness. For the very first tests, a
powerful feature of the SuperNEMO front-end electronics was exploited.
On each board,  it is possible to generate an electronic signal, with a
gate shape, which is injected into each signal channel. These pulses are
then transmitted through the coaxial signal cables to the PMT dividers,
where they are  then reflected by the RC circuit back to the
electronics. This feature allows us to test the whole chain of signal
transmission and its quality. We called these measurements {\it
reflectometry}.

\begin{figure}[htb]
\centering
\includegraphics[width=0.99\textwidth]{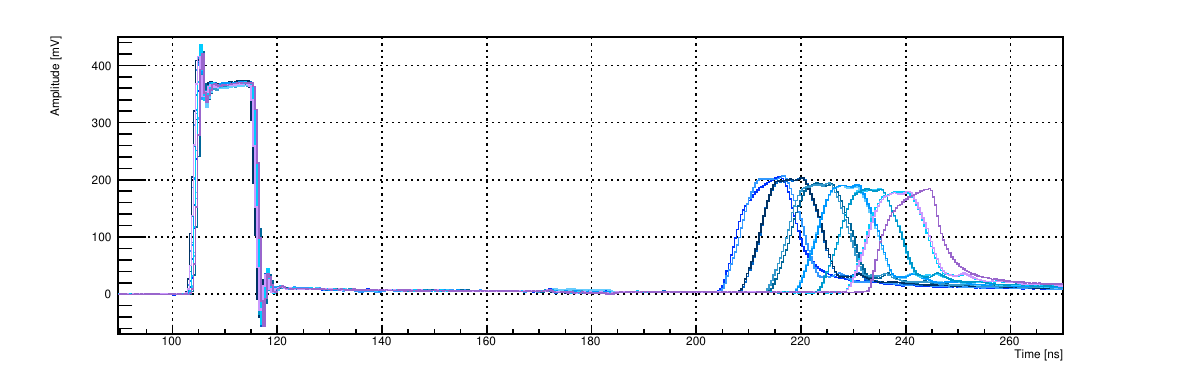}
\includegraphics[width=0.99\textwidth]{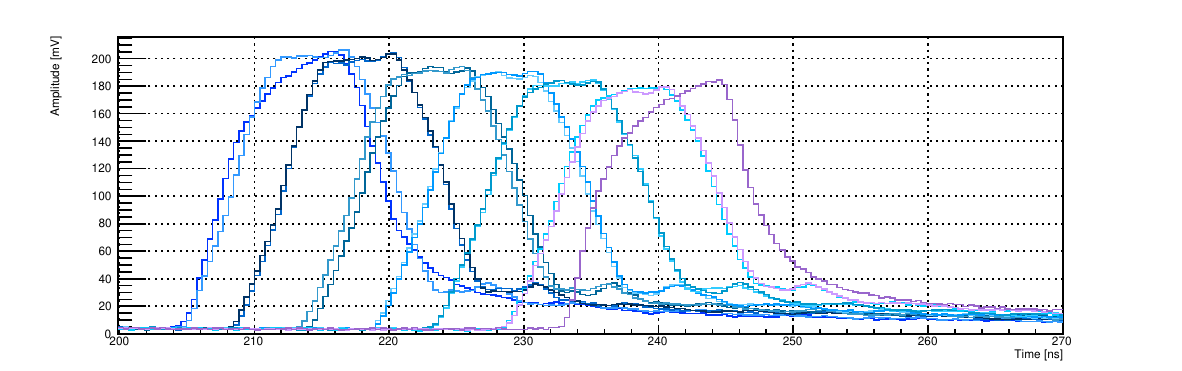}
\caption{\label{fig:reflecto-pulse} (top) Reflectometry example of the
averaged waveforms of the 13 channels of one front-end board. The
generated pulses can be seen around 105$\,$\si{\nano\second} and the
reflected pulses starting from 205$\,$\si{\nano\second}. The time shift
and the attenuation of reflected pulses are explained by increasing
cable lengths. (bottom) Zoom on the reflected pulses. The two wave-like
shapes (the first and last pulses) correspond to the top and bottom
5-inch PMTs, whereas the others are coming from 8-inch PMTs.}
\end{figure}

Thanks to the 400$\,$\si{\nano\second} length of the sampling window and
the maximum cable length of about 18 m, both the generated and the
reflected pulses can be digitized and stored in the same waveform. An
example of such a waveform for one FEB with 13 channels is shown in
figure~\ref{fig:reflecto-pulse}. We can see the first generated pulses
around 100$\,$\si{\nano\second} and the reflected pulses starting after
200$\,$\si{\nano\second}. Concerning the reflected pulses, we observe
three features:

\begin{itemize}
\setlength\itemsep{0em}
\item There is a few-nanosecond time shift between the reflected pulses.
This is due to the increasing cable length. The routing of the signal
cables is made up of rows of two, one above and one below the cable. The
cable lengths are separated by approximately 50 cm. This is why the
reflected pulses come in pairs, except for the 13\up{th}, which is alone.
\item The reflected pulse shapes vary with the increasing cable length.
We observe an attenuation in amplitude, which is accompanied by a
broadening of the pulses.
\item Two different shapes of reflected pulses can be seen,
corresponding to the 5- and 8-inch PMTs. The first and the last pulses,
corresponding to the 5-inch PMTs, present a wave-like shape. The others,
corresponding to the 8-inch PMTs, are closer to a gate signal. This can
be explained by the difference in the decoupling resistances on the PMT
dividers, which are 1 \si{\Mohm} and 10 \si{\kohm} for the 5-inch and
8-inch PMTs, respectively.
\end{itemize}

The analysis of the reflectometry data allowed us to perform a visual
inspection of the reflected pulses. Firstly, we needed to check if the
time position of each reflected pulse was correct. For example, we
detected issues at the patch panel when the pulses were reflected at a
shorter time than expected. We could also spot wrong cable lengths in
this way. Secondly, the shape of the reflected pulses allowed us to
investigate connection defects or damage to the connectors or the PMT
dividers. All these minor defects were easily corrected. The
reflectometry measurements provided an efficient quality check of the
calorimeter's signal cabling, and allowed us to have fully-operational
signal cabling.

\subsection{PMT signal verification}

\begin{figure}[!htb]
\centering
\includegraphics[width=\textwidth]{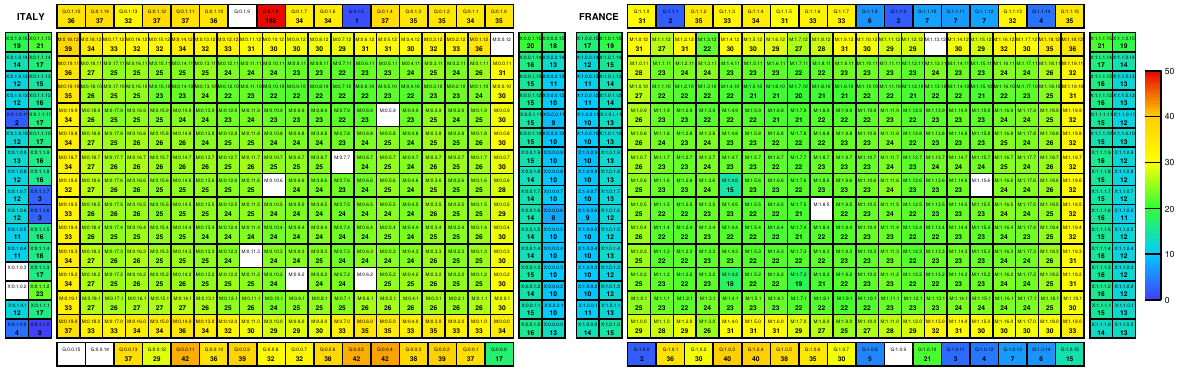}
\caption{Illustration of the functioning PMTs of the SuperNEMO
calorimeter, as of Summer 2023, without detector shielding. The picture
represents the count rate per minute (indicated by bold numbers and
color map) of $\gamma$-like events, for the two sides of the detector
(Italy and France). The white boxes correspond to disconnected cables at
the time of the measurement or dead PMTs. Some low-gain G-veto OMs also
appear at the top and bottom, but have been recovered later.}
\label{fig:pmt-maps}
\end{figure}

In order to continue with the tests of the calorimeter signal cabling,
we profited from the PMT signals. First, we switched on the PMTs one by
one and checked, through the acquisition, whether the correct channel
was triggering and showing PMT pulses. This test also allowed us to
validate the correspondence between HV and FEB channels. Secondly, a
visual inspection of the PMT pulses helped to find more problematic
channels. A few problematic channels revealed some broken PMTs. These
were confirmed by looking from the back of the PMT, where the normal
yellowish color of the PMT was gone. This means the photocathode had
disappeared because of a broken vacuum. This damage might have been
caused by a shock on the PMT, or by too much stress on the pins exiting
the glass. Since the PMTs were tested at LSM just before integration on
the wall, damage certainly occurred during the screwing-together of the
calobricks. A summary of the non-functioning PMTs can be seen in
figure~\ref{fig:pmt-maps}. In total, 16 out of 712 PMTs are
non-operational, which represents a dead-channel rate around 2\%.

%-----------------------------%

\section{Time calibration of the calorimeter}
\label{sec:time-calibration}

The goal of the time calibration is to compensate for  all relative
timing differences between any two channels of the SuperNEMO
calorimeter. These time differences can be due to different cable
lengths; to light collection in the scintillator; or to the PMT transit
time, which depends on the HV. To a minor extent, it can also be due to
the collection of signals in the backplane of the front-end electronics
crates.

\subsection{Reflectometry time measurements}

The amount of cables inside the anti-radon tent has been minimized in
order to reduce the radioactive background. Thus, the PMT cables have
been cut to the exact lengths to reach the PMTs from the patch panel.
This causes time delays between the different channels of the segmented
calorimeter, which need to be calibrated. In SuperNEMO, the timing
performance is of major importance. In particular, time coincidences are
needed to select the two electrons originating from the source foil, and
to reject crossing electrons, which are external background events
produced by $\gamma$-rays. This can be done thanks to a time-of-flight
analysis.

\begin{figure}[!htb]
\centering
\includegraphics[width=0.98\textwidth]{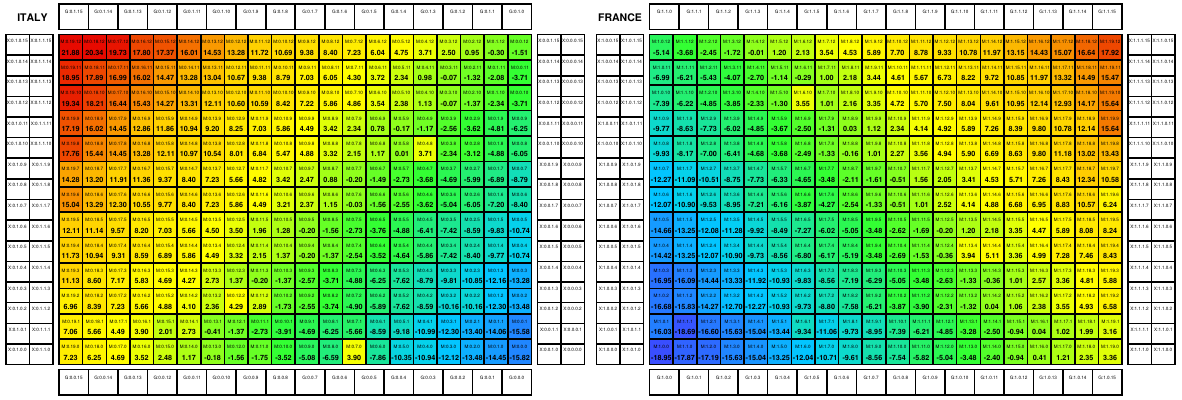}
\caption{Half-time difference measurements in ns (indicated by bold
numbers and color map) between the generated and reflected pulses in
reflectometry runs, for the two calorimeter main walls
\cite{girardcarillo:tel-03357651}. Some cables have slightly different
lengths than expected ; they appear in different colors to their
neighbours. The tiny text references correspond to the identification of
the OMs.}
\label{fig:reflecto-time-differences}
\end{figure}

The reflectometry technique, presented in
section~\ref{ssec:reflectometry-test}, also provides an efficient way to
measure the time differences between all of the channels. The analysis
of the waveforms was exploited to measure the time difference between
the generated and reflected pulses for each channel, which would
correspond to twice the time delay for PMT signals. All the time
differences were measured for the 712 channels thanks to this technique.
Figure~\ref{fig:reflecto-time-differences} shows the result for the main
wall OMs. We can observe the regular increase in cable lengths for the
two main walls, from the patch-panel corner in the bottom to the
opposite top corner, which results in an increase in the time differences.

\subsection{Time calibration with \up{60}Co source}

A \up{60}Co source is very well suited for performing a  time alignment
of several detectors. It emits two $\gamma$-rays within a very short
time coincidence ($<$ 1 ps), with 1.17 and 1.33 MeV energies. Comparing
the time measurements of these two gammas detected in two different OMs
would integrate all the processes (electronics, instrumental and
physics) that could produce time delay or jitter in the time
measurement. Following the reflectometry measurements, this method was
used to perform a complete time calibration of the SuperNEMO calorimeter
and, subsequently, to study its time resolution. The source was placed
about one meter away from a main calorimeter wall, with nine positions
used for each main wall. The measured time $t^\text{meas}_i$ of a
calorimeter hit $i$ can be written as :
\begin{equation}
t^\text{meas}_i = \text{ToF}_i + \kappa_i
\label{eqn:time-abs}
\end{equation}
where $\kappa_i$ is the time calibration constant to be determined for
each OM \textit{i} and $\text{ToF}_i$ is the time-of-flight of the gamma
particle from the source to the hit calorimeter \textit{$i$}. The time
difference between two OMs, $\Delta t_{ij} = t^\text{meas}_j -
t^\text{meas}_i$, will provide the difference between the calibration
constants $\kappa$ by :
\begin{equation}
\kappa_j - \kappa_i = \text{ToF}_i - \text{ToF}_j + \Delta t_{ij}
\end{equation}
The difference of time of flight $\text{ToF}_j - \text{ToF}_i$ is easily calculated
from the distance traveled by each $\gamma$-particle from the position
of the \up{60}Co source to the interaction point in the calorimeter.
This interaction point is unknown and we assume the $\gamma$-particle
interacted in the middle of the scintillator. The time difference
distributions $\Delta t_{ij}$ are then fitted by a Gaussian function.
The fitted mean value is used to determine the calibration constants,
and the fitted standard deviation to determine the uncertainty. As the
coincident-gamma rate decreases when the distance between two OMs
increases, the calibration procedure has to be limited to neighboring
OMs then iterated. Therefore the error on the calibration constant for
an OM at the edge of the wall is higher.

\begin{figure}[!htb]
\centering
\includegraphics[width=0.49\textwidth]{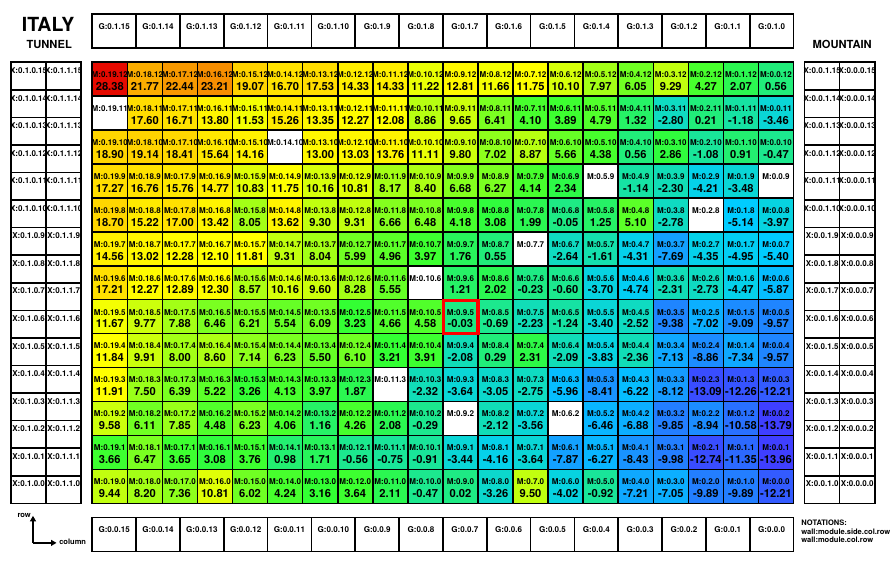}
\includegraphics[width=0.49\textwidth]{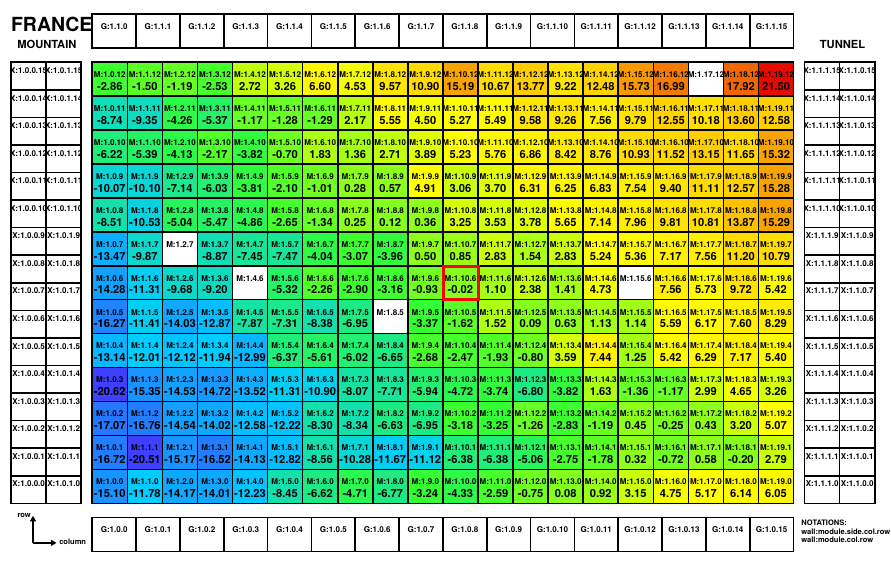}
\caption{Time difference in ns (indicated by bold numbers and color map)
to the central OM of each wall, with the \up{60}Co data, determining the
relative calibration constants $\kappa_i$ for the two calorimeter main
walls. The two red squares represent the two central reference OMs. The
white boxes correspond to dead PMTs or issues in calibrating a specific
OM with these data \cite{hoballah:tel-03865332}.}
\label{fig:co60-time-differences}
\end{figure}

By convention, we define a reference OM with $\kappa_\text{ref} = 0$ ns
, to which all other OMs will be relatively calibrated. Each M-Wall has
one reference OM located at its center. The result of this process for
all the M-wall OMs is presented in
figure~\ref{fig:co60-time-differences}. The values have also been
computed for the X-wall and G-veto blocks, but are not presented here.

We recognize the pattern dominated by the cable lengths, as in
figure~\ref{fig:reflecto-time-differences}. It confirms that the other
effects are lower-order corrections to the time delay. The statistical
uncertainties on these calibration constants present a peaked
distribution around 30$\,$ps for each of the M-walls, with a standard
deviation around 20$\,$ps. However, there are several uncertainty values
extending up to $\sim$100$\,$ps, for OMs at the edge of the wall, which
will require more statistics or further studies. Some systematic studies
were performed to test the validity of the results by varying selection
criteria in the analysis, and the results were found to be stable
\cite{hoballah:tel-03865332}.

In order to verify these results, a comparison with the reflectometry
measurements was performed. Good correlation was observed since the
difference between the calibration constants is below one nanosecond.
The dispersion is, however, larger since a standard deviation of about
2$\,$ns has been found. The reasons for these differences are
understood, because the $^{60}$Co measurements also include time jitters
(difference in transit times, transit time spread) introduced by the PMTs.

After applying the calibration constants to the time measurements, a
final verification consisted of checking the time-of-flight (ToF)
alignment of the two $\gamma$'s emitted from the \up{60}Co source. The
distribution of the ToF differences between the two OMs is well centered
at zero \cite{hoballah:tel-03865332}. The standard deviations are about 120$\,$\si{\pico\second} for
the M-wall. The choice of a reference OM with the best time resolution
in the M-wall could improve the precision of the calibration for all the
OMs. Nevertheless, these results are quite impressive given the status
of the detector in this early commissioning.

\subsection{Determination of the calorimeter time resolution with
\up{60}Co source}

After the alignment of the calorimeter OMs, we can measure the time
resolution of each OM. This will be decisive for the quality of the ToF
analysis, which is based on a $\chi^2$ comparison with a formula of the
type $\chi^2 = (\Delta t)^2 / \sigma_t^2$, where $\Delta t$ is the time
difference between the registration of signals from two OMs and
$\sigma_t$ the coincidence time resolution of the calorimeter, see for
example \cite{boursette:tel-01885274}. The objective is to be comparable
to NEMO-3, for which $\sigma_t$ = 660$\,$\si{\pico\second} has been
measured with 1$\,$MeV $\gamma$-rays \cite{augier:tel-00011894}, despite
the fact that SuperNEMO scintillators blocks are larger.

\begin{figure}[!htb]
\centering
\includegraphics[width=0.49\textwidth]{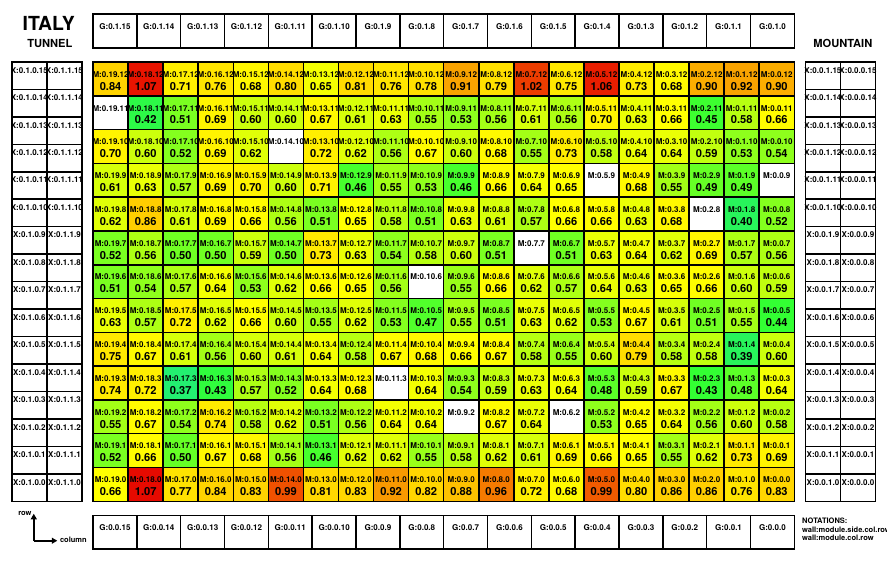}
\includegraphics[width=0.49\textwidth]{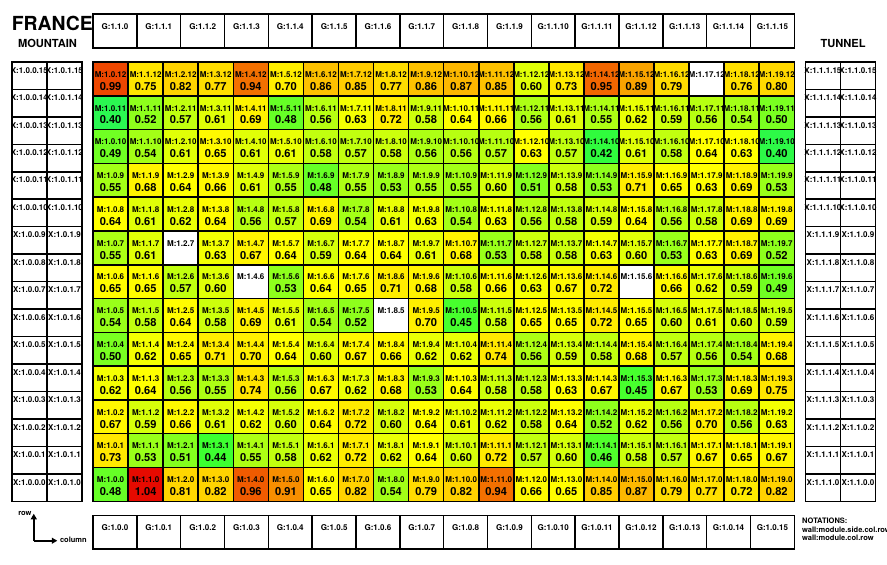}
\caption{Time resolutions $\sigma_i$ in ns (indicated by bold numbers
and color map) of each OM of the two calorimeter main walls obtained
with the \up{60}Co calibration source \cite{hoballah:tel-03865332}. The
white boxes correspond to dead PMTs or issues in calibrating a specific
OM with these data.}
\label{fig:co60-time-resolution}
\end{figure}

In order to determine the individual time resolution of each OM, a new
method has been developed. The uncertainties on the time difference
measurements, denoted $\sigma_{ij}$, represent a combination of the time
resolution $\sigma_i$ for the OM $i$ and $\sigma_j$ for the OM $j$:
\begin{equation}
\sigma_{ij}^2 = \frac{\sigma_i^2}{E_{ij}} + \frac{\sigma_j^2}{E_{ji}}
\label{eqn:sigma-time}
\end{equation}
where $E_{ij}$ corresponds to the energy measured by the OM $i$ in the
events in coincidence with the OM $j$, and $E_{ji}$ the opposite.
Energies must be taken into account since the time precision depends on
the number of detected photons, and thus on the energy. In order to
determine the individual time resolutions $\sigma_i$, we can triangulate
between three OMs and use the measurements for the pairs $ij$, $jk$ and
$ik$. Solving the system of three equations allows us to determine the
three resolutions of interest. Repeating these calculations allows us to
get the individual resolutions for all the OMs of the SuperNEMO
calorimeter. The results of this method for the M-wall OMs are presented
in form of maps in figure~\ref{fig:co60-time-resolution}. The time
resolutions are quite uniform around 615 ps, except for the top and
bottom rows, which show poorer time resolution since the 5-inch PMTs of
these OMs collect less light.

\begin{figure}[!htb]
\centering
\includegraphics[width=0.85\textwidth]{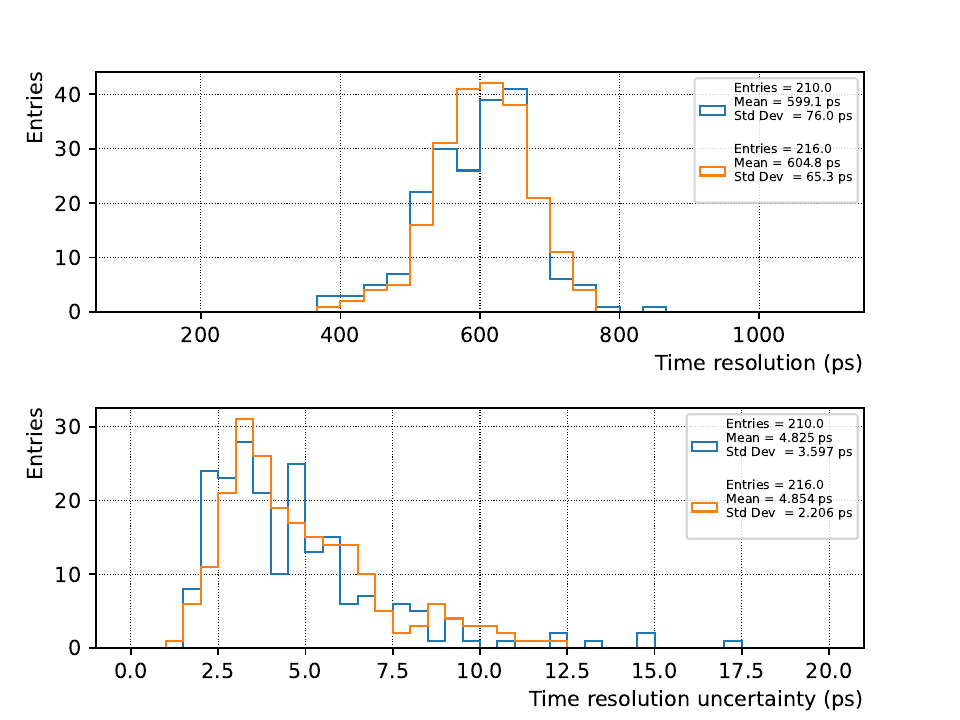}
\caption{Time resolution of the 8-inch OMs (top) with the corresponding
statistical uncertainties (bottom). The orange and blue histograms
correspond to the OMs of Italian-side and French-side M-walls respectively.
Reproduced from \cite{hoballah:tel-03865332}.}
\label{fig:co60-time-resolution-error}
\end{figure}

Similar maps were generated for the uncertainties on the time resolution
for each OM \cite{hoballah:tel-03865332}. These maps revealed larger
uncertainties at the edges of the detector, which could be improved with
higher collected statistics. The final one-dimensional distributions of
the time resolutions and the corresponding uncertainties are presented
in figure~\ref{fig:co60-time-resolution-error}.

In order to provide a global time resolution of the SuperNEMO
calorimeter, weighted averages have been computed for the 5- and 8-inch
PMTs. Systematic studies varying the analysis cuts were performed to
estimate the systematic uncertainties on these averaged resolutions. The
final result for the time resolution of the main walls of the SuperNEMO
calorimeter, extracted from the \up{60}Co data, is presented in
table~\ref{tab:sn-time-resolution}. Since the 8-inch PMTs collect more
scintillation photons, the time resolution is better for these OMs
compared to the 5-inch.

\begin{table}[!htb]
\centering
\renewcommand{\arraystretch}{1.3}
\begin{tabular}{|l|c|c|}
\cline{2-3}
\multicolumn{1}{c|}{} & 8-inch PMTs & 5-inch PMTs\\
\hline
Italian-side M-wall & $614 \pm 2\ \text{(stat)}\ ^{+64}_{-~1}\ \text{(syst)}$ ps & $828 \pm 5\ \text{(stat)}\ ^{+101}_{-~~1}\ \text{(syst)}$ ps \\
\hline
French-side M-wall & $619 \pm 2\ \text{(stat)}\ ^{+49}_{-~4}\ \text{(syst)}$ ps & $814 \pm 6\ \text{(stat)}\ ^{+73}_{-~1}\ \text{(syst)}$ ps \\
\hline
\end{tabular}
\caption{Weighted average of the individual time resolutions of the
SuperNEMO OMs from the calorimeter M-walls extracted from the \up{60}Co data
\cite{hoballah:tel-03865332}.}
\label{tab:sn-time-resolution}
\end{table}

The final time resolution of about 615$\,$ps with $\gamma$-particles is
slightly better than the 660$\,$ps measured for NEMO-3
\cite{augier:tel-00011894}. This result suggests that we may reach a
better value than 250$\,$ps for 1-MeV electrons in SuperNEMO, compared
to NEMO-3, which is the final objective of the time analysis. The time
resolutions are significantly better for electrons since they interact
within the first millimeters of the scintillator's front face. We can
also expect to improve the time analysis of SuperNEMO, since this work
was done at an early stage of the detector commissioning, with limited
statistics.

%-----------------------------%

\section{Energy calibration of the calorimeter}
\label{sec:energy-calibration}

Precisely measuring the energy of the two electrons from a DBD event is of major importance for SuperNEMO, to search for
$0\nu\beta\beta$ at the end of the $2\nu\beta\beta$ energy spectrum.
Having the best possible energy resolution is essential for this
separation. The energy calibration and its stability over time is one of
the most important tasks for the experiment. The energy calibration of
the SuperNEMO calorimeter will be mostly performed by regular deployment
of the \up{207}Bi sources, between the $\beta\beta$ source foils in the plane of the source frame. The deployment system
was already active at the time of the calorimeter commissioning, but as
the activity of the sources is very low ($A < 150$ Bq
\cite{SuperNEMO:2021hqx}), it is impossible to identify the \up{207}Bi
conversion electrons without the tracking detector and without
shielding. Thus, another method has been used to perform the energy
calibration of the calorimeter, which was inspired by \cite{Loaiza:2017mpb}.

\subsection{Energy calibration using \boldmath$\gamma$-rays}

Given the importance of the energy measurement, studies of the SuperNEMO
calorimeter's energy response  began as soon as was possible
\cite{pin:tel-03149593}, despite the inability to reconstruct electrons.
Since the shielding had not yet been installed, we had an opportunity to
attempt the first energy calibration of SuperNEMO's calorimeter using
the external $\gamma$'s emitted by the rock of the laboratory. This
$\gamma$~flux at LSM is dominated by decays of the  \up{40}K, \up{214}Bi
and \up{208}Tl isotopes. Since they have different energies, it is
possible to detect several Compton edges on the total energy spectrum in
each calorimeter block, providing several energy-calibration points at a
time. There are almost ten $\gamma$-rays of significant intensity to
consider in the \up{214}Bi decays, which cannot be identified
individually, given the energy resolution of the calorimeter, but its
total spectrum can still be used. The 1460$\,$keV of \up{40}K and the
2615$\,$keV of \up{208}Tl are, on the other hand, easily identified.

In order to fit both the position and the rate of all these
$\gamma$-rays in the energy spectrum of each OM, simulations of the
three isotopes, with initial vertex positions at the walls of the LSM,
were performed. This enabled the creation of {\it probability density
functions} (PDF), which have been used to fit the measured spectra,
using the RooFit library provided by ROOT \cite{root}.

Having several energies tested at the same time is also very important
for SuperNEMO because several non-linearity effects have to be taken
into account. These effects have been studied by the SuperNEMO
collaboration, using an electron spectrometer beam
\cite{Marquet:2015ima} and precise optical simulations
\cite{huber:tel-01628463}. Three main effects can produce
non-linearities in the scintillator blocks: inhomogeneities based on the location in the scintillator relative to the PMT, Birks quenching of the scintillation light
\cite{birks1964} and the production of Cherenkov light by  higher-energy
charged particles. The geometrical light-collection effect dominates,
because of the large volume of the scintillator blocks. It can lead to a
50\% increase for light produced just in front of the photocathode compared to the middle of the scintillator's entrance face, and
a 10\% loss in the corners. Birks' law produces a few-percent increase
above 1 MeV\footnote{1 MeV has been chosen as the reference for the
normalization of the relative effects.} but, most importantly, a loss of
10\% at 200 keV. Similarly, the Cherenkov effect can produce a 2\%
increase in the collected energy above 1 MeV, and a  2\% decrease below.
These effects were applied to the simulated PDFs before using them to
fit the measured energy spectra of the OMs \cite{aguerre-2023}.

\begin{figure}[!htb]
\centering
\includegraphics[height=0.3\textheight]{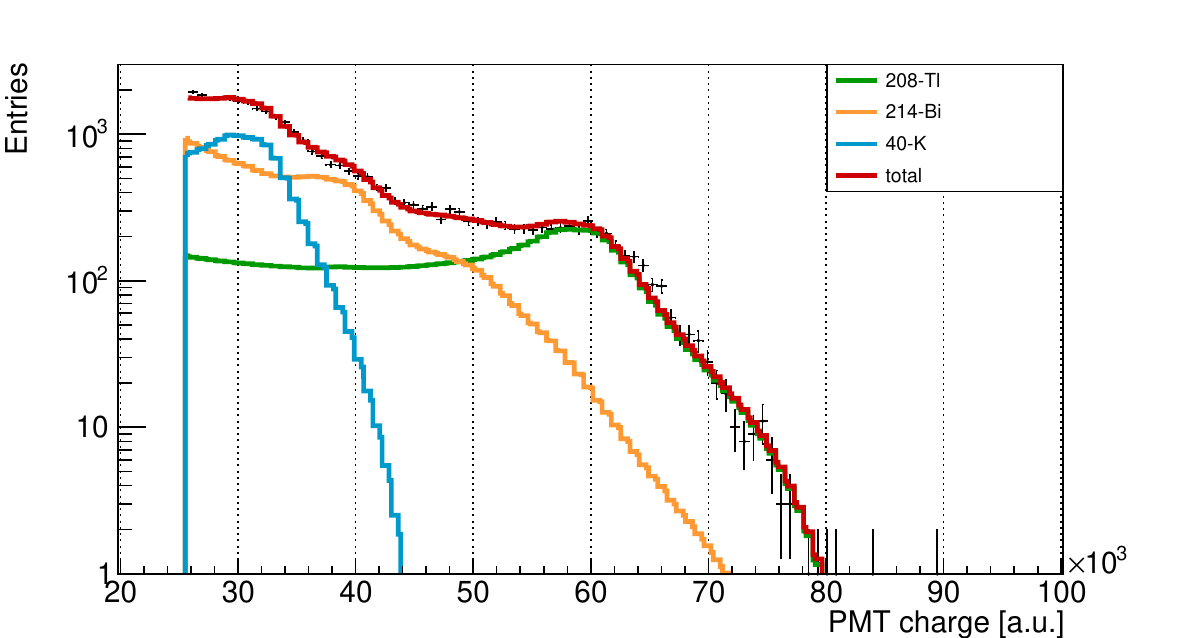}
\caption{Example of the result of the energy calibration of the OM M-wall
1.9.9 of the SuperNEMO calorimeter. The contributions from the three
external background radionuclides are represented in cyan for \up{40}K, in
orange for \up{214}Bi and in green for \up{208}Tl. The red line
represents the total reconstructed spectrum that gives the best
agreement to the data points, shown in black \cite{aguerre-2023}.}
\label{fig:lsm-calibration-spectra}
\end{figure}

The energy spectrum of each OM can then be fitted to determine the five
needed parameters: one calibration constant, one energy resolution
(res) of the form $\sigma_E = \text{res} \times \sqrt{E}$, and one activity
for each of the three isotopes. The two first parameters are scanned in
a defined range and the activities are adjusted by the fit. A
statistical $\chi^2$ test is computed at each step of the scans and the
best calibration parameters are determined by fitting the $\Delta
\chi^2$ curve, to find the minimum. An example of the calibration result
for one M-wall OM is presented in figure~\ref{fig:lsm-calibration-spectra}.

\subsection{Alignment of the calorimeter OMs}

\begin{figure}[!htb]
\centering
\includegraphics[width=\textwidth]{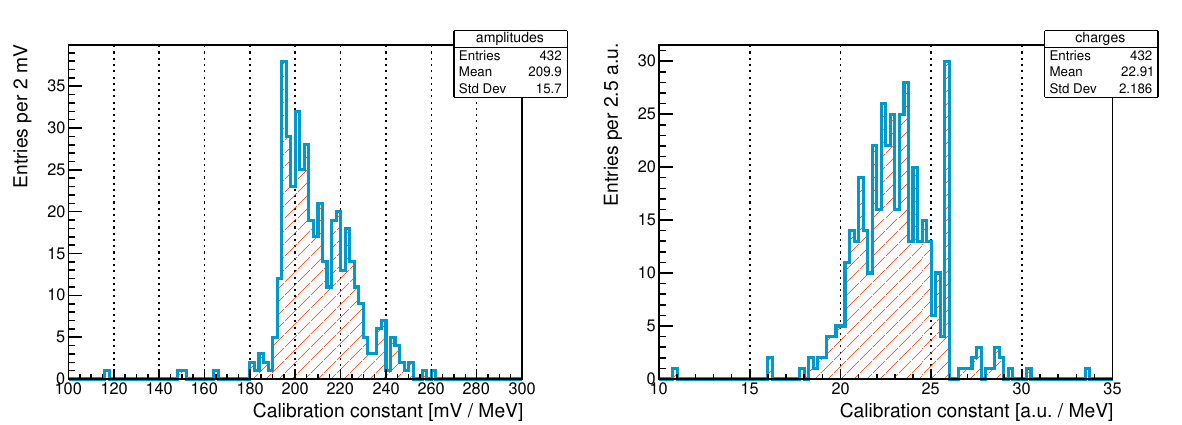}
\caption{Distribution of the calibration constants of the M-wall OMs of
the SuperNEMO calorimeter in amplitude (left) and charge
(right)\cite{aguerre-2023}. The calibration constants are aligned here
relative to the PMT pulse amplitude, because this is the variable of
interest for the trigger threshold.}
\label{fig:result-energy-calibration}
\end{figure}

Almost all the optical modules have been calibrated using this new
method. Figure~\ref{fig:result-energy-calibration} presents the current
status of the energy calibration for the 8-inch OMs of the M-wall, on
the amplitude and the charge spectra. A good alignment, at a level of
about 7.5\%, has been obtained on the PMT pulse amplitudes (standard deviation divided by the mean of the histogram in figure~\ref{fig:result-energy-calibration} left), which is the
relevant parameter to align trigger thresholds. Moreover, the amplitude
histogram shows a low-amplitude cutoff, which is due to the requirement
of having a minimal gain of 190$\,$mV/MeV. This value was optimised in order to ensure a sufficient coverage of an energy range from 150 keV up to around 12 MeV, given the $\pm$1.25$\,$V dynamic range of the electronics. Nevertheless, the charge of the
pulse is a better estimate of the measured energy, compared to the signal amplitude, and will be used in
the event reconstruction.

This SuperNEMO energy calibration method has proven to be very powerful
in determining the calibration constants of the OMs. However, no clear relationship was found between energy resolution and a minimal $\chi^2$ for the fit of activities; a step-wise increase in the energy-resolution parameter in the model resulted in large fluctuations in the best-fit $\chi^2$ values, with no observable trend. A dedicated study will
be conducted on this crucial parameter for SuperNEMO, using precise
electron tracking and energy measurement.\\

This work has also been used to measure the flux of the ambient
$\gamma$-rays at LSM, whose result will be the subject of another
publication. It will give a reference spectrum for the simulation of the
external background in the SuperNEMO experiment, as well as for other
experiments located in the LSM.

%-----------------------------%

\section*{Conclusions and outlook}
\label{sec:conclusions-and-outlook}
\addcontentsline{toc}{section}{Conclusions and outlook}

This article is the first SuperNEMO publication presenting an analysis
of data from the SuperNEMO Demonstrator. The data were acquired at an
early stage of the experiment, when the tracking detector was not yet
commissioned, and the detector shielding was not yet installed. However,
it was already possible to commission the SuperNEMO calorimeter with
$\gamma$-particles in this configuration. We have used either
calibration sources or the ambient-$\gamma$ background of the LSM.

We have first reported a validation of the functioning of the 712 OMs
with their signal- and high-voltage cabling, despite a few broken PMTs.
The reflectometry method used for this validation also allowed us to
measure the time delays between all the OMs, which will be important to
detect coincidence events, like double-beta-decay.

Secondly, we have presented the use of two almost-coincident
$\gamma$-rays emitted by a \up{60}Co source, to calibrate, in time, all
the calorimeter OMs. This second time alignment allowed us to account
for all the effects that could produce relative delays between two OMs and achieved a precision of about 120~\si{\pico\second}.
This analysis also allowed us to extract the time resolution for
$\gamma$-particles, which is around 615$\,$ps for the OMs of the
SuperNEMO calorimeter M-wall.

Finally, we have performed the energy calibration of the SuperNEMO
calorimeter with the ambient-$\gamma$ background of the LSM, which allowed an alignment of about 7\.5%. A novel
fitting method using the three natural
radionuclides, \up{40}K, \up{208}Tl and \up{214}Bi, has been
implemented. It also allowed us to incorporate optical corrections to
reproduce the measured spectra. This improved the understanding of the
detection of $\gamma$-particles in the SuperNEMO scintillator blocks.

At this stage all the functioning OMs have been aligned at in time and in energy. The calorimeter can thus be used for physics data taking, which is about to
start. The tracking detector has now been commissioned and the shielding
integration is almost complete. This configuration will allow us to
determine crucial calorimeter parameters, such as energy and time
resolutions, with electron tracks.

%-----------------------------%

\acknowledgments

The authors would like to thank the staff of the LSM (Laboratoire Souterrain de Modane), Université Grenoble Alpes, Univ. Savoie Mont Blanc, Grenoble INP, CNRS, IN2P3, LPSC/LSM, 38026 Grenoble, France, for their technical assistance in assembling and operating the detector. 

We also would like to thank the CC-IN2P3 computing center, Centre de Calcul de l’Institut National de Physique Nucléaire et de Physique des Particules, CNRS, IN2P3, 69627 Villeurbanne, France, providing all the resources for data transfer, storage processing and analysis.

Finally, we acknowledge support by the grants agencies of the CNRS/IN2P3 in France, Czech Republic, NSF in the U.S.A., RFBR in Russia, Slovakia and STFC in the UK.

Finally, we acknowledge support by CNRS/IN2P3 in France, the MEYS of the Czech Republic (Contract Number LM2023063), NRFU in Ukraine (Grant No. 2023.03/0213), NSF in the USA, Slovak Research and Development Agency (APVV-15-0576, APVV-21-0377) and STFC in the UK.

%-------------
% bibliography

\end{document}